\documentclass[aps,pra,preprint,showpacs,floats,preprintnumbers,amsmath,amssymb,superscriptaddress]{revtex4}
\usepackage{epsfig}
\usepackage{graphicx}
\usepackage{amsmath}
\usepackage{CJK}

\def\be{\beta}

\def\fr{\frac}
\def\be{\begin{equation}}
\def\ee{\end{equation}}
\def\bea{\begin{eqnarray}}
\def\eea{\end{eqnarray}}
\def\bes{\begin{subequations}}
\def\ees{\end{subequations}}
\begin{document}

\begin{CJK*}{GBK}{song}

\title{Dynamics of shock waves in a superfluid unitary Fermi gas}

\author{Wen Wen}
\email{wenwen0emma@163.com}
\affiliation{Department of Mathematics and Physics, Hohai University, Changzhou 213022, China}
\author{Tiankun Shui}
\author{Yafei Shan}
\affiliation{College of Mechanical and Electronic Engineering, Hohai University, Changzhou
213022, China}
\author{Changping Zhu}
\affiliation{Jiangsu Key Laboratory of Power Transmission and Distribution Equipment
Technology, Changzhou Key Laboratory of Sensor Networks and Environmental
Sensing, Hohai University, Changzhou 213022, China}

\begin{abstract}
We study the formation and dynamics of shock waves initiated by a repulsive potential in a superfluid
unitary Fermi gas by using the order-parameter equation. In the theoretical framework, the regularization
process of shock waves mediated by the quantum pressure term is purely dispersive. Our results show
good agreement with the experiment of Joseph {\it et al}. [Phys. Rev. Lett. {\bf 106}, 150401 (2011)].
We reveal that the boxlike-shaped density peak observed in the experiment consists of many vortex rings
due to the transverse instability of the dispersive shock wave. In addition, we study the transition from a
sound wave to subsonic shock waves by increasing the strength of the repulsive potential and show a strong
qualitative change in the propagation speed of the wavefronts. In the relatively small strength regime,
the speed decreases below the sound speed with increasing the strength as a scaling behavior, while in the
large regime the speed remains almost unchanged, which is found to be the same expansion speed of the
proliferation of the vortex rings.

%\noindent{\it Keywords}:  unitary Fermi gas, shock wave, propagation speed, transverse instability

\end{abstract}

\pacs{03.75.Kk, 03.75.Lm, 03.75.Ss}

\maketitle

\section{Introduction}
The unitary Fermi gas is a system of fermions with an infinite
two-body scattering length, which has remarkable universal
properties and has connections with areas as diverse as nuclear
physics and high-$T_c$ superconductivity \cite{zwe2012}. The
unprecedented experimental sophistication reached recently in the
unitary Fermi gas has opened a unique opportunity for systematic
studies of elementary excitations in a strongly interacting Fermi
system \cite{ket2008,all2012}. Experiments have observed collective
modes \cite{kin2004,bar2004,tey2013} and sound waves
\cite{jos2007,sid2013}, nonlinear topological structures including
vortices \cite{zwi2005} and dark solitons \cite{yef2013,ku2014}, and
another specific nonlinear phenomenon, i.e. shock waves
\cite{jos2011}. Shock wave generated in a classical fluid is
characterized by a well-defined shock front across which, there is a
dissipation from the effect of viscosity to avoid the onset of a
gradient catastrophe \cite{whi1974,lan1987}. In contrast to viscous
shock wave, shock wave in a Bose-Einstein condensate
is mediated by dispersion. Such dispersive shock wave associated
with a fast oscillating wave-train trailing has been predicted
theoretically \cite{kul2003,dam2004,kam2004,vic2004,hof2006}, and
then observed in various experiments \cite{cha2008,
dut2001,mep2009}.

In the recent experiment performed by the
Thomas's group \cite{jos2011}, the formation and propagation of shock waves in a unitary Fermi gas
were studied by colliding between two clouds. They reproduced the experimental data well by
hydrodynamic equations with a phenomenological viscosity term. On the other hand, Bulgac {\it et al.}
\cite{bul2012} and Salasnich {\it et al.} \cite{anc2012,sal2011,anc2013,lsa2013} independently argued that
the observed fermionic shock front is dominated not by dissipation but rather dispersion, because at zero
temperature the bulk viscosity vanishes \cite{son2007,cao2011} and the shear viscosity is at a minimum
\cite{cao2011,wal2012}. Since the structures of shock wave fronts are too fine to be imaged in the
typical Fermi gas experiments, the regularization mechanism is difficult to be directly revealed through
density measurements. In order to determine the appropriate regularization mechanism, Lowman and Hoefer
\cite{low2013} subsequently proposed to measure the propagating speed of shock waves which is regularization
dependence \cite{hof2006}. It is noticed that the above mentioned references
\cite{anc2012,sal2011,anc2013,lsa2013,low2013} mainly present the results of one dimensional (1D)
dynamics of fermonic shock waves by integrating over the transverse coordinate. There is a lack of discussion on
the transversal effect on shock wave dynamics, however, which is always present in Fermi gas experiments
and of considerable practical interest.

In this work, we use the order-parameter equation to perform a detailed three-dimensional (3D) numerical
simulation of all stages in the experiment of fermionic shock waves \cite{jos2011} and compare to experiment
with good agreement. We find that a dispersive shock wave forms, which possess an expanding oscillatory
wavetrain with a large amplitude. Due to the transverse instability, the wavetrain decays into a large number of
vortex rings in a short time. We interpret that the observed boxlike-shaped density peak in the
experiment actually originates from the proliferation of the vortex rings. Furthermore, we study the mechanism
of transition from sound wave propagation to shock wave dynamics and present how the wavefront speed depends
on the strength of the repulsive potential which creates it. For the smallest strength creating a linear wave,
the propagation speed is in good agreement with the theoretically predicted sound speed. As increasing the
strength in the relatively weak regime, we find that the speed decreases below the sound speed.
In this regime we derive an analytical expression for the speed as a function of the strength
from an effective 1D model, which is also confirmed by the full 3D numerical simulation exactly. As the strength
increases towards the moderate value, the transversal dynamics takes an effect on the properties of shock waves,
and the speed deviates the scaling behavior predicted by the 1D model. As increasing the strength in the
sufficiently large regime, shock waves are formed by colliding two spatially separated clouds, and the propagation
speeds of the wavefronts are found to be unchanged. We understand the speed independent on the strength
as the expansion speed of the vortex rings, which are originated from the decay of the dispersive shock waves
due to the transverse instability.

This paper is organized as follows: In Sec. II, we solve numerically the order-parameter equation to
study the formation and dynamics of fermionic shock waves in the realistic system. In Sec. III,
we analytically and numerically study the transition from sound wave propagation to subsonic shock wave
dynamics by calculating the propagation speeds of the wavefronts. Last section gives a conclusion.

\section{Comparison with experiment}
\subsection{Theoretical description}
We consider an ultracold Fermi gas at zero temperature, in which
fermionic atoms have two spin states with equal number and all atoms are paired
in the superfluid state. The dynamic behaviors of Fermi superfluids can be
described by the following time dependent order-parameter equation
\cite{sal2008,sal12008,adh2008,adh12008,wen2008}
\be\label{gg}
i\hbar\fr{\partial \Psi_s}{\partial
t}=\left[-\fr{\hbar^2\nabla^2}{4m}+2V_{\rm ext}({\bf r})+2\mu (n)\right]\Psi_s,
\ee
where $\Psi_s$ is the order parameter of fermionic atomic pairs in
the condensed state and the atomic density is given by $n=2|\Psi_s|^2$,
$m$ is the mass of atom, and $V_{{\rm ext}} ({\bf r})$ is the external potential.
In the unitary limit the s-wave scattering length becomes infinity and
no characteristic length is set, so the equation of state has the form
\cite{ast2004,bul2006}
\begin{equation}
\mu(n)=\xi\frac{\hbar^2}{2m}(3\pi^2)^{2/3}n^{2/3},
\end{equation}
where the Bertsch parameter $\xi=0.38$ is tuned to fit the experimental results
\cite{nav2010,ku2012}.

This complex order parameter can be specified by $\Psi_s=\sqrt{\frac{n}{2}}e^{i\Phi_s}$,
where $\Phi_s$ is the phase of fermionic pair condensates. By introducing the superfluid
velocity given by the gradient of the phase ${\bf v}=(\hbar/2m)\nabla \Phi_s$, from the
order-parameter equation one can deduce the quantum hydrodynamics equations
\bes\label{hydrody}
\bea & &\label{hydrody1}\frac{\partial n}{\partial t}+\nabla\cdot(n{\bf
v})=0, \\
& & \label{hydrody2}m\frac{\partial {\bf v}}{\partial t}+\nabla
\left[\frac{1}{2}mv^2+\mu (n)+V_{{\rm ext}}({\bf r})-\frac{1}{4}\frac{\hbar^2}{2m}
\frac{\nabla^2\sqrt{n}}{\sqrt{n}}\right]=0,
\eea
\ees
which includes a gradient correction of von-Weizs$\ddot{a}$cker energy form
\cite{zub2009,zub12009,and2010} (i.e. quantum pressure term)
\begin{equation} \label{weit}
\lambda\frac{\hbar^2}{2m}\frac{\nabla^2\sqrt{n}}{\sqrt{n}}
\end{equation}
with $\lambda=1/4$. In absence of the quantum pressure term, Eqs.(\ref{hydrody}) reduce to
the classical hydrodynamic equations \cite{lan1987,men2002}.

In order to prevent the density gradient catastrophe and model the shock wave formation,
Ref. \!\cite{jos2011} introduces a dissipative viscosity term $\nu\partial_z(n\partial_z v)/n$
into the normalized 1D form of the classical hydrodynamic equations,
where $\nu=10\hbar/m$ is fitted to the experimental results.
However, it is shown that Fermi gases in the unitary regime
have a vanishing viscosity at zero temperature \cite{son2007,cao2011,wal2012}.
For the quantum hydrodynamics equations (\ref{hydrody}), an alternative regularization mechanism
can be naturally proposed, without the phenomenological viscosity term. The gradient term (\ref{weit})
that takes into account corrections to the kinetic energy due to spatial variations in the density
of the system, plays a role of a pure dispersive effect for the formation of shock waves.
There has been a lot of discussion about the value of the coefficient $\lambda$\;
\cite{and2010,kim2004,sal12008}. The quantum pressure term depending explicitly on the reduced
Planck constant $\hbar$ gives quantization condition, and the value of $\lambda$ has profound
consequences on dynamics of superfluid unitary Fermi gases \cite{anc2012,anc2013,lsa2013}.

Our calculations are based on the order-parameter equation, which models only the dynamics of
superfluid components. To further include the mechanism for superfluid relaxation, such as pair-breaking,
superfluid-normal transition, and various photon processes, one must resort to time-dependent
density functional theory (DFT) which provides a microscopic description \cite{bulg2013,bulg2011}.
Recently based on DFT, many different dynamical processes have been investigated, including formations
of vortex lattices by stirring \cite{bulg2011}, crossing and reconnection of two vortex lines \cite{bulg2011}, dynamics
of shock waves \cite{bul2012}, oscillations of a vortex ring \cite{bul2014}, and conversion between vortex rings
and vortex lines due to a breaking of the axial symmetry of traps \cite{wla2014}, some of which can not be
obtained from the order-parameter equation. While the order-parameter equation fails to model the pair-breaking
mechanism and normal components, the order-parameter equation depending on a single collective wave function is
significantly easier to solve analytically and numerically than DFT, which requires to solve hundreds of thousands
of wavefunctions by means of supercomputing resources. In addition, Forbes and Sharma \cite{forb2013} have presented
a comparison between the dynamics of superfluid unitary Fermi gases using DFT and the order-parameter equation (they call
the extended Thomas-Fermi model), and demonstrated that the order-parameter equation is a good description for
low-frequency dynamics of unitary Fermi gases.

\subsection{Numerical results}
%===========================fig1===============================%
\begin{figure}
\includegraphics[scale=0.3]{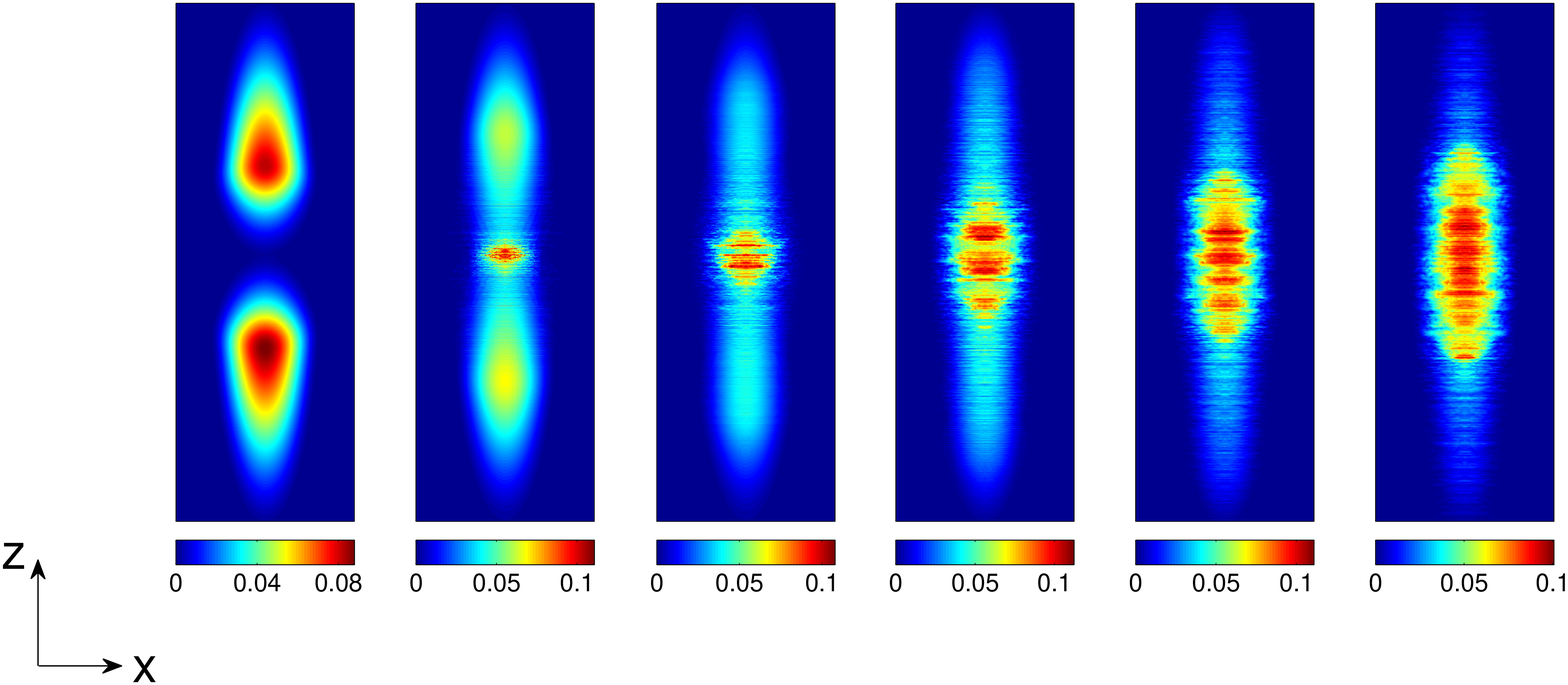}
\caption{\footnotesize{(Color online) Integrated atomic density profiles
(in units of $10^{-3}/\mu m^2$ per particle) after expansion for 1.5\;{\rm ms} after evolving
in the trap for different times: from left to right $t=$ 0\;{\rm ms}, 2\;{\rm ms},
4\;{\rm ms}, 6\;{\rm ms}, 8\;{\rm ms}, and 10\;{\rm ms}. Here $x$ and $z$ correspond to
the coordinates in $\mu m$, with $x\in(-100,100)$ and $z\in(-210,210)$.}}
\end{figure}
%===========================fig1===============================%
%%%%%%%
We solve the order-parameter equation numerically \cite{adh12010}
to reproduce as closely as possible the experimental condition. The experimental
procedure including three steps is described as follows \cite{jos2011} :
(1) The unitary Fermi gas containing a total of $N=2\times 10^5$ atoms
is initially confined in a cigar-shaped harmonic trap\;\;
$V_{\rm ho}({\bf r})=\frac{1}{2}m[\omega^2_{\perp}(x^2+y^2)+\omega^2_zz^2]$
with $\omega_{\perp}=2\pi\times437\;{\rm Hz}$ and $\omega_z=2\pi\times27.7\;{\rm Hz}$,
and bisected by a repulsive potential \;$V_{\rm rep}(z)=V_0\exp{[-(z-z_0)^2/\sigma_z^2]}$
with strength $V_0=12.7\;\mu K$, width $\sigma_z=21.2\;{\rm \mu m}$ and offset $z_0=-5\;{\rm \mu m}$.
(2) The repulsive potential $V_{\rm rep}$ is then suddenly turned off, allowing for the
two separated parts of the cloud to collide with each other in the harmonic trap for a given hold time $t$.
(3) The absorption images are finally taken after an additional $1.5\;{\rm ms}$ expansion, during which
the harmonic potential in the $r$-direction
is extinguished and the $z$-direction frequency is changed to $\omega_z=2\pi\times20.4\;{\rm Hz}$.

The creation of initial density perturbations undergoing
shock wave dynamics achieved experimentally can be classified into two ways
\cite{dam2004,cha2008,sal2011,mep2009}: suddenly turning off a repulsive potential produces a
dip in the density, which splits into two negative perturbations propagating in opposite
directions; alternatively, raising suddenly a repulsive potential
results in  a density bump, which splits into two positive perturbations. Due to the density-dependent
speed that is regions of high density move with a faster local velocity than regions of low density,
as they travel the steepenings of leading edges of positive density waves and
of trailing edges of negative ones are apparent. Therefore, according to the experimental condition,
two negative wave packets propagating in opposite directions are actually observed.

%%%%%%
%===========================fig2===============================%
\begin{figure}
\includegraphics[scale=0.35]{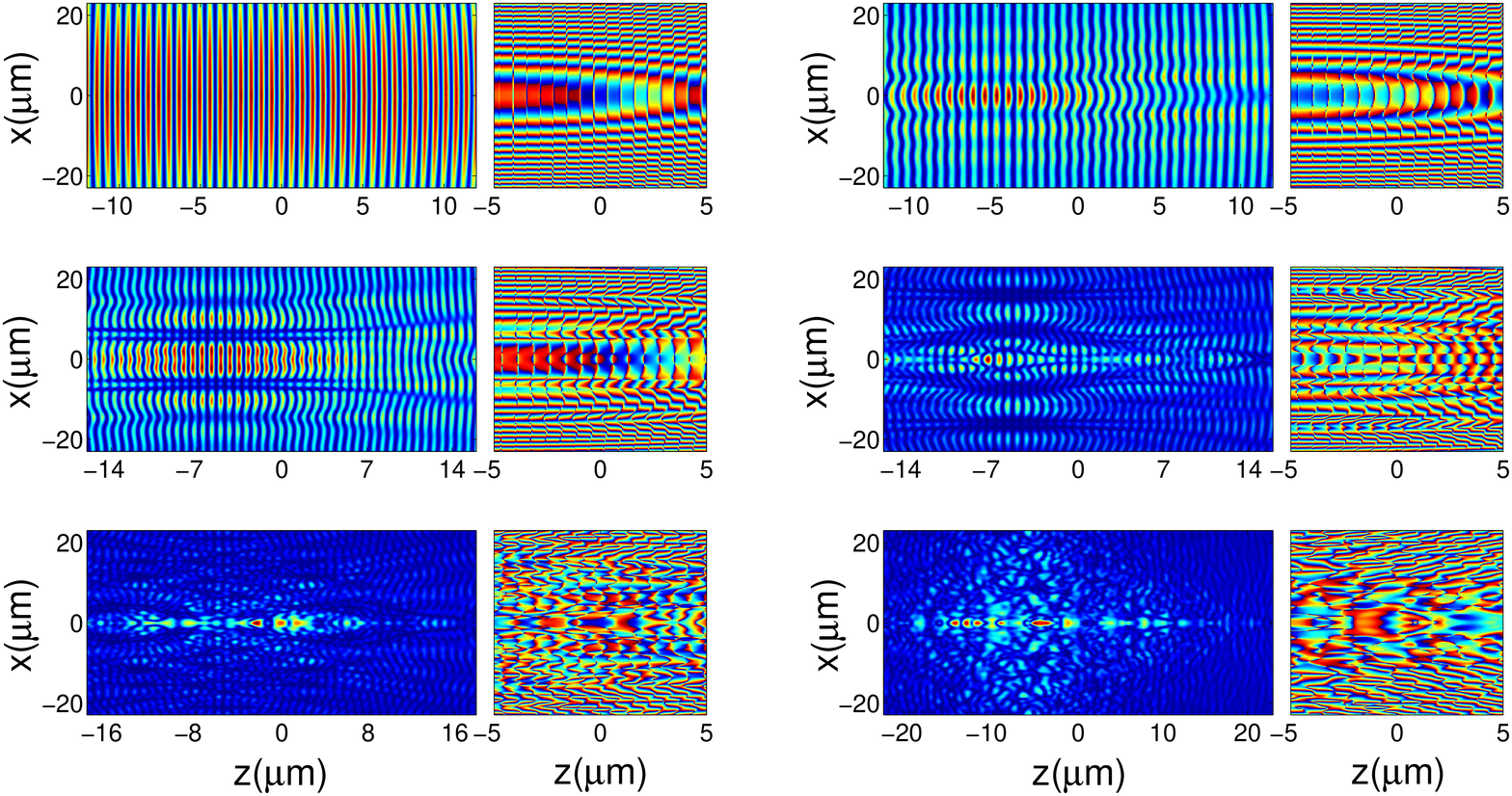}
\caption{\footnotesize{(Color online) Closeup of the atomic density slices $n(x,0,z)$ (left panels)
and the corresponding phase profiles $\Phi_s(x,0,z)$ (right panels) after expansion for 1.5\;{\rm ms}
after evolving in the trap for different times,
from top left to bottom right: $t = 0.7\;{\rm ms}$, $ 1.1\;{\rm ms}$, $1.2\;{\rm ms}$, $1.3\;{\rm ms}$,
$1.5\;{\rm ms}$ and $2.0\;{\rm ms}$. The color corresponds to the phase: $\Phi_s = 0$ through $2\pi$
is represented by the sequence blue-green-red-blue. }}
\end{figure}
%===========================fig2===============================%
%%%%%%%
In Figure 1, we show the numerical results for  column atomic densities $\int \!dy\; n(x,y,z,t)$
by integrating along the $y$ axis at different hold times $t$. It is seen that the presence of
the repulsive potential leads to two clearly separated parts. After the potential is rapidly turned off,
the two clouds expand and collide at the center of the trap, and a pronounced bulge of higher atom
density forms. As the density bulge spreads out from the center of the trap,
its front (back) edge self-steepens, which can be identified as the formation of a shock wave.
Notice that the observed front and back edges actually correspond to the trailing edges of
two negative waves propagating in opposite directions, respectively. The result on the feature
of the box-like structure is found to be in good agreement with the experimental observation
[see Fig. 1 of Ref.\;\cite{jos2011}]. However, clear soliton trains in the wake of the density
bulge which are indication of the dispersive mechanism are not been found, as that in the experiment.

To better visualize the complicated density distributions at the
center of the system and find the dispersive effect, in Figure 2 we present zoomed-in views for the
atomic density slices by the plane $y=0$, that is $n(x,y=0,z,t)$, and the corresponding phase profiles
$\Phi_s(x,0,z)$. A closer look to the center region
reveals that a soliton train forms after evolving in the trap at $t=0.7$\:{\rm ms}. Subsequently the
soliton train is curved as it develops. The curving structure is not symmetric about the $z=0$ axis,
due to the offset $z_0=-5\;{\rm \mu m}$ of the initial repulsive potential. When $t=1.3$\;{\rm ms},
the transverse instability causes the quick decay of the soliton train into the nucleation of vortex rings
\cite{bul2014,cet2013,mun2014,hoe2009,wen2013,fed2000}, evidenced by the phase profiles.
After 2.0\;{\rm ms}, a large number of vortex
rings spread over the entire extent of the system. The formation of vortex rings in an initial short
time is shown to be difficult to detect unambiguously in the experiment that are integrated along
the line of sight \cite{jos2011,cha2008}. While the phenomena are similar to those found in atomic
condensates \cite{cha2008}, they are first revealed in a superfluid unitary Fermi gas.
Recently based on DFT \cite{bul2012}, Bulgac {\it et al.} studied the shock wave
formed in the unitary Fermi gases trapped by a two-dimensional harmonic potential.
Because the particle number of the calculated system is too small, no transverse instability is found and
the dynamics is characterized by the formation of a soliton train that eventually fills the entire superfluid.
Our numerical simulation can be consistent with this (not shown here) if we use small number particle.

\section{transition from a sound wave to subsonic shock waves}

Now we study the mechanism of the transition from sound wave propagation to shock wave dynamics by
increasing the strength of the repulsive potential and calculate the propagation speeds of the wavefronts.
Before presenting the results from a full 3D numerical calculation, we reduce Eqs.\;(\ref{hydrody}) into
an effective 1D model, derive an analytical solution for the propagating speed that is only valid for
small perturbations, and make a comparison with the numerical results.

\subsection{Analytical derivation}
We consider a cylindrically symmetric external potential composed of the harmonic potential in the
transverse direction and an axial potential: $V_{{\rm ext}}({\bf r},t)=\fr{1}{2}m\omega^2_{\perp}r^2+V_1(z,t)$,
with $r=\sqrt{x^2+y^2}$. The atomic density and superfluid velocity are then expressed as
\;\cite{low2013}
\begin{equation} \label{fac}
n({\bf r},t)=n_{\perp}[r;n_1(z,t)]n_1(z,t)\;\;\;\; {\rm and}\; \;\;\; {\bf v}({\bf r},t)=v_1(z,t).
\end{equation}
The atomic density satisfies the normalization conditions: $2\pi\int dr \: rn_{\perp}=1$ and $\int \:dz n_1=N$.
We assume sufficiently tight transverse confinement so that its dynamics is neglected. Under
the Thomas-Fermi (TF) approximation, the density distribution in the equilibrium is given by
\begin{equation} \label{tradensity}
n_{\perp}[r;n_1]=n_{\perp}(0)(1-\fr{r^2}{R_{\perp}^2})^{3/2},\;\;\;\;\; R^2_{\perp}=\fr{2\mu_{\perp}}
{m\omega^2_{\perp}}.
\end{equation}
By the normalization condition $2\pi\int^{R_{\perp}}_0 dr rn_{\perp}$=1,
the center atomic density and chemical potential are then given by
\begin{equation} \label{chemcen}
n_{\perp}(0)=(\fr{2m\mu_{\perp}}{\xi\hbar^2})^{3/2}(6\pi^2n_1)^{-1},\;\;\;\; \mu_{\perp}=\bar{\mu}n^{2/5}_1=
(\fr{15\pi m\omega^{2}_{\perp}}{4})^{2/5}(\fr{\xi\hbar^2}{2m})^{3/5}n^{2/5}_1,
\end{equation}
respectively. Substituting the above ansatz into Eqs.\;(\ref{hydrody}) and integrating over the
transverse coordinate, one can arrive at the effective 1D hydrodynamic equations
\bes\label{hydro1D}
\bea & &\label{hydro1D1}\frac{\partial n_1}{\partial t}+\fr{\partial}{\partial z}(n_1v_1)=0, \\
& & \label{hydro1D2}m\frac{\partial {v_z}}{\partial t}+\fr{\partial}{\partial z}
\left[\frac{1}{2}mv_z^2+\bar{\mu}n^{\fr{2}{5}}_1+V_1\right]=0,
\eea
\ees
with $\bar{\mu}$ defined in Eq.\;(\ref{chemcen}). Notice that the power of the nonlinear term for the density
has changed from $2/3$ to $2/5$ as we pass from
the 3D decription Eqs.\;(\ref{hydrody}) to the effective 1D model Eqs.\;(\ref{hydro1D}).
In additon, we assume that the spatial scale of the density perturbation is larger
than the healing length and neglect the gradient correction (i.e. quantum pressure) term.

In the case of the axial harmonic potential $V_1=\fr{1}{2}m\omega^2_zz^2$, one has the ground-state solution
of Eq.\;(\ref{hydro1D2})
\begin{equation}
n_1(z)=n_1(0)(1-\fr{z^2}{R^2_z})^{5/2},  \;\;\;\; R^2_z=\fr{2\mu_G}{m\omega^2_z},
\end{equation}
where $\mu_G$ is the ground-state chemical potential fixed by the normalized condition $\int^{R_z}_{-R_z} dzn_1=N$.
It is easy to get the explicit expressions
\begin{equation} \label{chemcen1}
n_1(0)=(\mu_G/\bar{\mu})^{5/2},\;\;\;\;\; \mu_G=\sqrt{\xi}E_F,
\end{equation}
with the Fermi energy $E_F=\hbar(3N\omega^2_{\perp}\omega_z)^{1/3}$.
After linearization of Eqs.\;(\ref{hydro1D}), one can obtain the local speed of sound calculated at the center
of the trap \cite{cap2006,wen2010}
\begin{equation}\label{sp}
c^2_s=\frac{2\bar{\mu}}{5m}n^{2/5}_1(0)=\fr{1}{5}\xi^{\fr{1}{2}}v^2_F
\end{equation}
with the Fermi velocity $v_F=\sqrt{2E_F/m}$ . We find that such local sound speed
$c_s=\sqrt{\xi^{1/2}/5}v_F$
is smaller than the sound speed $c_{\rm ho}=\sqrt{\xi/3}v_F$ of a homogeneous system \cite{hei2006}
by a factor of $\sqrt{3/5\xi^{1/2}}=0.987$, which is resulted from the suppression of the transverse confinement.

We perform the ansatz on
the density $n_1(z,t)=n_1(0)\rho(z,t)$, which implies a homogenous density background characterized by the central density.
By substituting it and Eq.\;(\ref{sp}) into Eqs.\;(\ref{hydro1D}), the effective 1D hydrodynamic equations are written as
\bes\label{shydro}
\bea & &\label{shydro1}\frac{\partial \rho}{\partial t}+\rho\fr{\partial v_z}{\partial z}+v_z\fr{\partial \rho}{\partial z}=0, \\
& & \label{shydro2}\frac{\partial {v_z}}{\partial t}+v_z\fr{\partial v_z}{\partial z}+\fr{c^2_{ls}}{\rho}\fr{\partial \rho}{\partial z}=0,
\eea
\ees
with
\begin{equation}
c_{ls}(\rho)=c_s\rho^{\fr{1}{5}}.
\end{equation}
In order to find wave solutions of Eqs.\;(\ref{shydro}), it is assumed that the velocity $v_z$ depends explicitly the density $\rho$ \cite{sal2011}.
In this case one can put $\partial v_z/\partial t=\frac{d v_z}{d \rho}\frac{\partial \rho}{\partial t}$, $\partial v_z/\partial z=\frac{d v_z}{d \rho}
\frac{\partial \rho}{\partial z}$, and reduce Eqs.\;(\ref{shydro}) to a hyperbolic equation
\begin{equation}\label{hyperbo}
\fr{\partial \rho}{\partial t}+C(\rho)\fr{\partial \rho}{\partial z}=0.
\end{equation}
It is easy to find the exact traveling wave solution of the hyperbolic equation: $\rho=f(z,C(\rho(z,t))t)$, where $f$ is an arbitrary function
and\;$C(\rho)=v_z+c^2_{ls}(dv_z/d \rho)^{-1}/\rho$ is the local speed of propagation.  By a requirement that
far from the perturbation the density is equal to one and the velocity field is zero, the local speed is then
given as $C(\rho)=\pm c_s(6\rho^{1/5}-5)$,
where the sign $\pm$ determines a direction of propagation.
%===========================fig3===============================%
\begin{figure}
\includegraphics[scale=0.2]{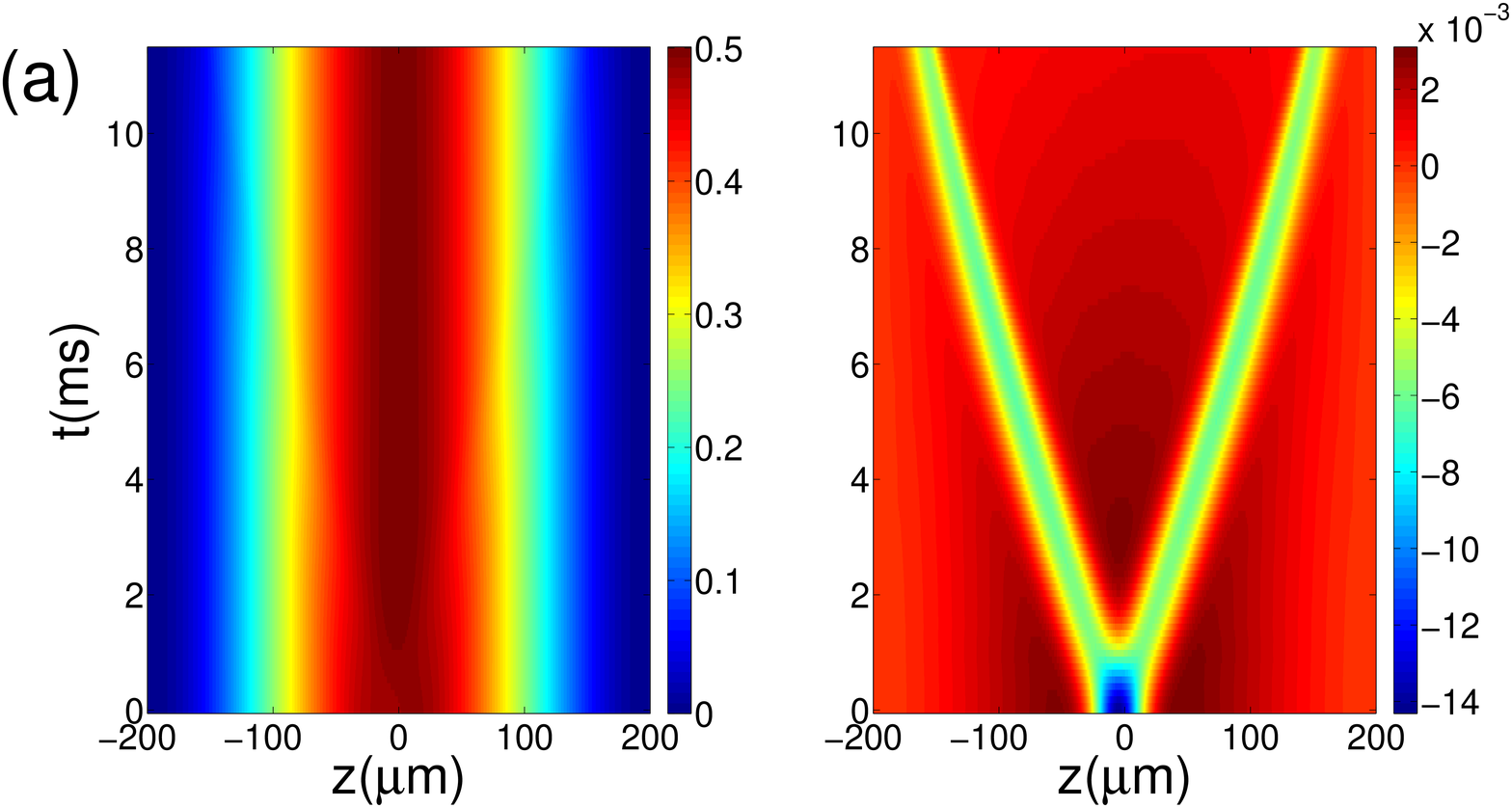}
\includegraphics[scale=0.2]{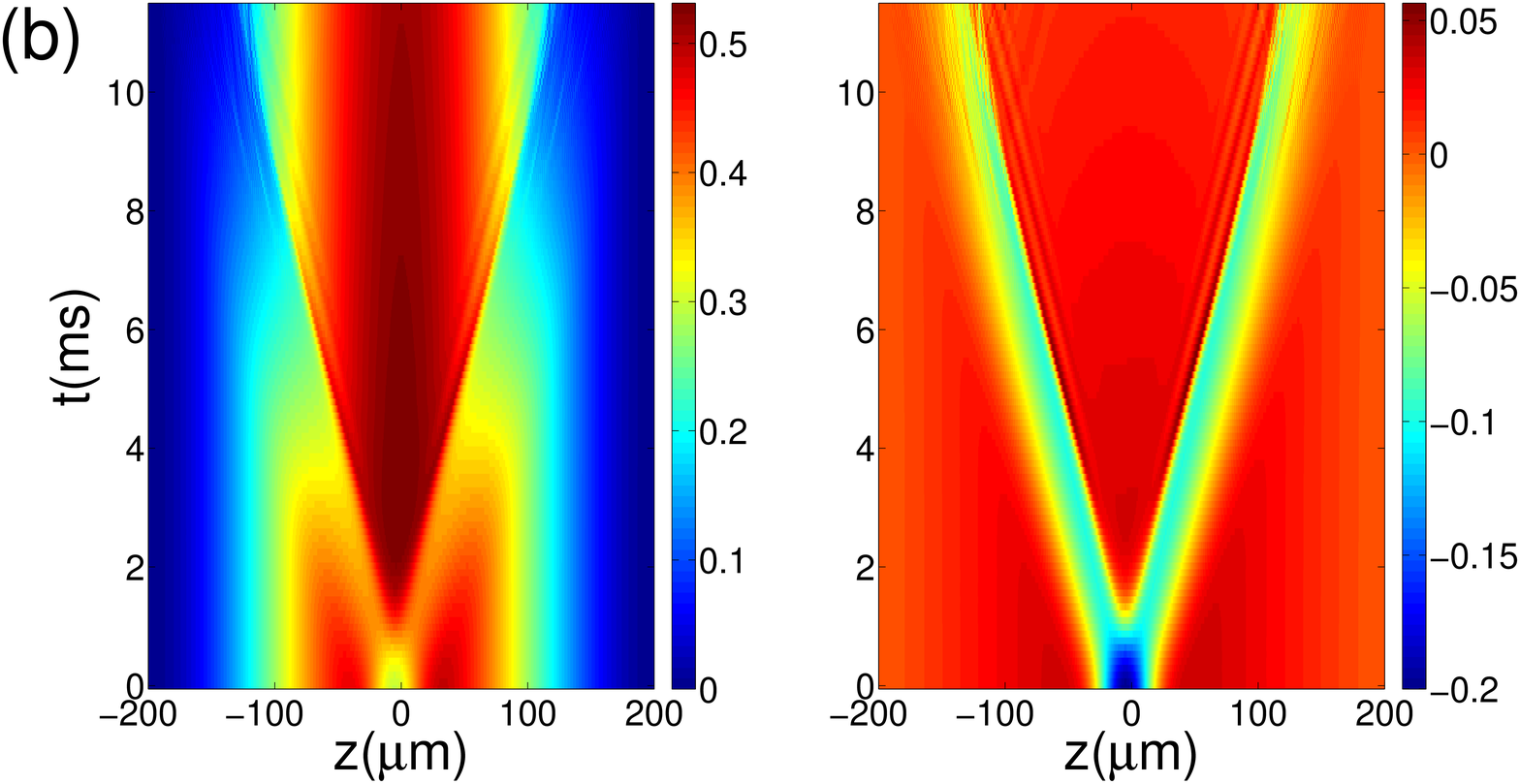}
\caption{\footnotesize{ (Color online) Space-time evolutions
in the harmonic trap of the atomic
density for relatively weak repulsive potential
strengthes $|A|=V_0/\mu_G$ with the chemical potential $\mu_G=0.435\mu K$.
The atomic density integrated along the transverse direction is in units of $10^{-2}/\mu m$
per particle. Panels (a) and (b) correspond to $|A|=0.0153$ and $0.23$, respectively.
Left: the actual density. Right: the normalized density (the actual density minus the ground-state density). }}
\end{figure}
%===========================fig3===============================%
%%%%%%%
The repulsive potential produces an initial density
$\rho(z,0)=(1+A\exp({-z^2/\sigma^2}))^{5/2}$, where the magnitude of $A$ is proportional to the strength, i.e. $|A|\varpropto V_0$.
After the repulsive potential is suddenly turned off or switched on, the Gaussion shaped perturbation breaks into two separate parts with
half-amplitude of the initial perturbation moving in the opposite directions \cite{sal2011,dam2004,dam12004}.
The extrema of the traveling wave has a constant amplitude $1+\eta$, and its speed is given by
\begin{equation}\label{sshock}
c(\eta)=c_s[6(1+\eta)^{1/5}-5],
\end{equation}
where the wave amplitude is determined by the strength of the repulsive potential through the relation
$\eta=[(1+A)^{5/2}-1]/2$ \cite{dam12004}. It is seen that the speed of shock waves explicitly depends on initial
conditions. In the limit $A\sim0$, the speed reduces to the sound speed $c_s$. For
$A<0$ ($A>0$) corresponding to negative (positive) perturbations, one has subsonic
(supersonic) waves.

%===========================fig4===============================%
\begin{figure}
\includegraphics[scale=0.2]{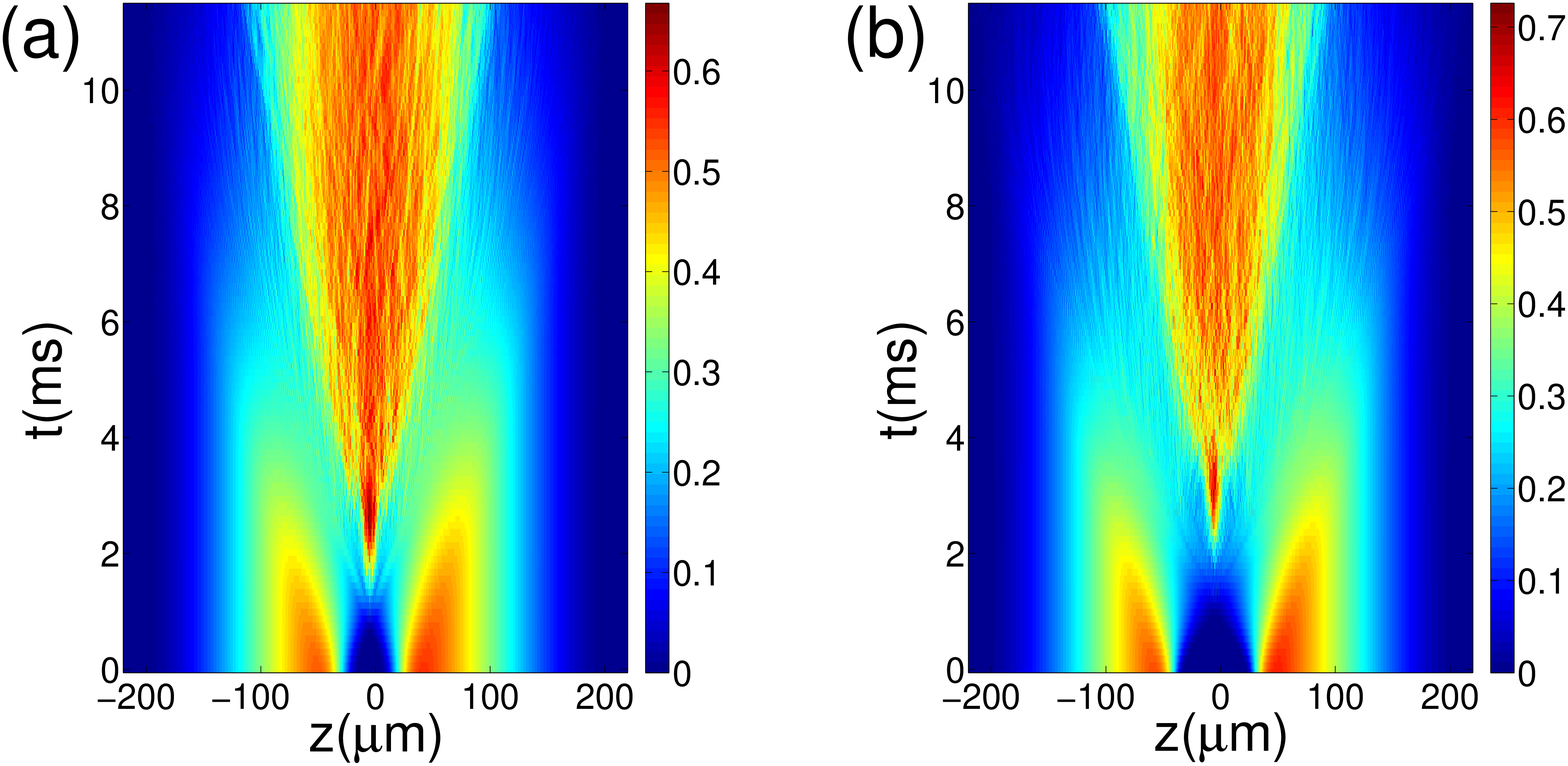}
\includegraphics[scale=0.2]{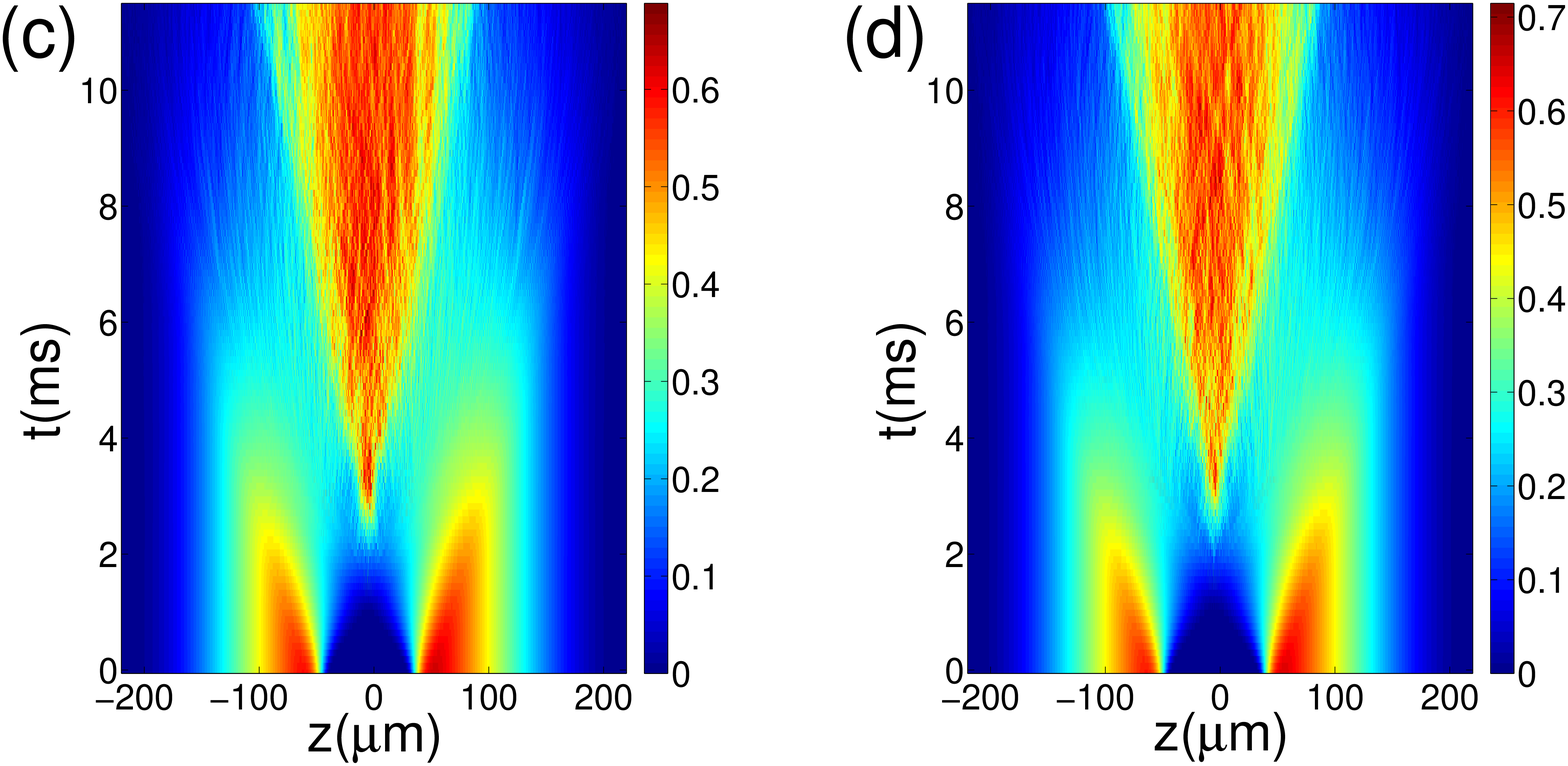}
\caption{\footnotesize{ (Color online) Space-time evolutions in the
harmonic trap of the atomic density for repulsive potential strengthes
$|A|>1$. The atomic density integrated along
the transverse direction is in units of $10^{-2}/\mu m$ per particle.
Panels (a), (b) and (c) correspond to
$|A|=1.22$, $6.11$, and $14.6$, respectively.
Panel (d) shows the experimental case $|A|=29.2$. }}
\end{figure}
%===========================fig4===============================%
%%%%%%%

In the above analytical discussion,  we assumed that the transverse density is approximated by the TF ground-state,
which means that the dynamics in the transverse direction is frozen and no instability occurs as studied in
Refs.\;\cite{sal2011,low2013,dam2004}. Salasnich obtained the propagation speed of the shock waves proportional to
$[4(1+\eta)^{1/3}-3]$ in a different configuration \cite{sal2011}, i.e. 3D homogenous unitary Fermi gas. Such expression
differs from our result only for the constants, because the integration of the transversal harmonic confinement leads
the power of the nonlinear term to change from 2/3 to 2/5. Therefore, one can predict the speed of the shock waves
in a homogeneous BEC in proportion to $[3(1+\eta)^{1/2}-2]$ \cite{dam2004}, due to the nonlinear power $2/2$.
Based on the effective 1D hydrodynamic equations (\ref{hydro1D}) incorporating the quantum pressure term (\ref{weit}),
Lowman and Hoefer obtained implicit relations for the speeds of supersonic shock waves \cite{low2013}. However, numerical simulation
in the following will show that the analytical result is only valid for relatively weak repulsive potentials.
As the strength of the potential increases, the dispersive shock wave is created with larger amplitude of
oscillating wave-trains. In a strongly interacting Fermi gas, they are easily subject to the transverse instability
that leads to the formation of vortex rings \cite{bul2014,cet2013,mun2014,wen2013}. Such decay process eventually takes
an significant effect on the dynamical properties of shock waves.

\subsection{numerical results}

In the following we numerically solve the order-parameter equation to study the formation and dynamics
of shock waves in the harmonic trap, which are initiated by different strengthes of the repulsive potential
under the experimental conditions. Due to the perturbations are created by suddenly turning off the repulsive potential,
they are negative and subsonic. The amplitude of the perturbation is then expressed by
$\eta=[(1-|A|)^{5/2}-1]/2$, where the magnitude $|A|=V_0/\mu_G$ is characterized by the ratio of strength of the
repulsive potential to chemical potential, and $\mu_G=0.435\;\mu K$ is obtained by substituting the experimental parameters
into Eq.\;(\ref{chemcen1}).

Figure 3 shows the space-time evolutions of the atomic density integrated along the transverse direction at relatively small
repulsive potential strengthes, and other parameters are the same as the experimental parameters.
Fig. \!\!3(a) shows the case of a very small perturbation $|A|=0.0153$.
However, in the left panel the propagation of the wave cannot be directly observed through the density profile. In order to
visualize the trajectory of the wave, we present the normalized density profile in the right panel, i.e. the actual density
subtracts the ground-state density. It is clear that the initially induced negative perturbation splits into two density dips,
which propagate outward at an almost constant speed and slow down when approaching the superfluid boundary due to the density
dependence of the speed. For such weak perturbation $|A|\sim0$, one might expect to excite linear wave. One can extract the
propagation speed from the center of the normalized density profile \cite{jos2007,sid2013}. The obtained
speed of $0.347v_F$ is consistent with the analytical prediction $c_s=\sqrt{\xi^{1/2}/5}v_F=0.351v_F$. With increasing $|A|=0.23$
in Fig. \!3(b), in the left panel we observe directly the formation of the shock wave from the density profile,
in which the shock front is represented as a sharp discontinuity of the two density lines forming a ``V" shape.
In addition, one can see that after a long time dynamics, the steepness of the shock wave front
is associated with oscillatory behaviors,
which are more obvious from the normalized density profile in the right panel.

%===========================fig5===============================%
\begin{figure}
\includegraphics[scale=0.3]{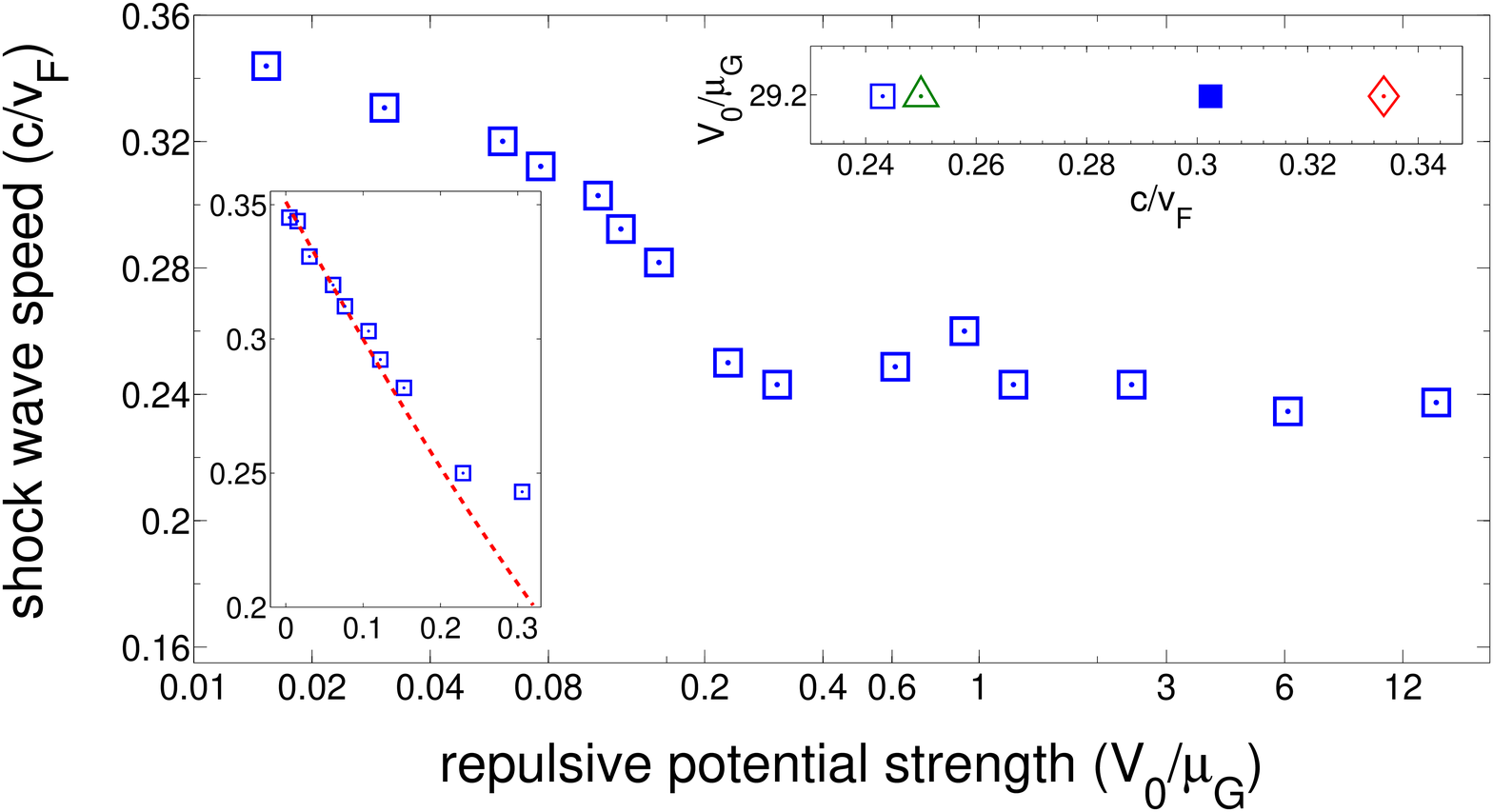}
\caption{\footnotesize{(Color online) Propagation speed of the shock wave fronts as a function of the strength
of the repulsive potential. Left insert: the comparison of the numerical simulations ($\Box$) and the analytical
prediction Eq.(\ref{sshock}) (dashed line). Right inset: the comparison of the results of the experiment ($\diamondsuit$)
\cite{jos2011}, the density functional theory ($\bigtriangleup$) \cite{bul2012}, and our results calculated for
the cases in the harmonic trap ($\Box$) and after an additional expansion (filled $\Box$).}}
\end{figure}
%===========================fig5===============================%
%%%%%%%

In Figure 4, we show the space-time evolutions of the atomic density initiated by large potential strengthes $|A| >1$,
which means that the repulsive potential results in two spatially separated clouds. Fig. \!4(a) corresponds to the case of $|A|=1.22$.
Different from the cases of initial small perturbations in Fig. \!3, the density oscillation behaivors begin to occur
in the overlap of the two clouds at the center of the system. For larger values of the strengthes
in Fig. \!\!4(b) ($|A|=6.11$) and Fig. \!4(c) ($|A|=14.6$), stronger oscillations of the densities in
more extensive regions are observed. But interestingly, from panels (a) to (d) we notice that the simulations present
very similar characteristic of the ``V" shape. Fig. \!4(d) shows the experimental situation $|A|=29.2$,
which is in agreement with the experimental results [see Fig. \!1 of Ref. \!\cite{bul2012}]. Noticed that compared
to the experimental observation in Fig. \!1 of Ref. \!\cite{bul2012}, which is on an expanded scale realized
by an additional $1.5\;{\rm ms}$ expansion after evolving in the harmonic trap ( i.e. all three steps described in
Sec. IIB are included), in Fig. \!3 and 4 we present the space-time evolutions in the harmonic trap without
the additional expansion (i.e. only the first two steps are included).

From Fig. \!3 and 4 we not only observe the transition from a sound wave to shock waves, but we also extract the propagation
speeds of the shock wave fronts by calculating the slopes of the two lines in the ``V" shape. The propagation speed of the shock
waves as a function of the repulsive potential strength is plotted in Figure 5. Two approximate regimes can be distinguished:
In the relatively small strength regime, the propagation speed decreases below the sound speed as the strength
increases.  In the left inset we compare the numerical results ($\Box$) to the analytical prediction Eq.\;(\ref{sshock}) (dashed line).
One can find that the analytical expression agrees very well with the
numerical results in the very small strength regime ($|A| < 0.2$). In this regime,
the transversal dynamics has a negligible effect, so that Eq.\;(\ref{sshock}) is an accurate expression for the speed of the shock waves
derived from the effective 1D model. Specifically, one can see that for $|A| >0.3$ the speed presents a
significant deviation from the analytical prediction, which indicates that the 1D model is not adequate in the description of the 3D
system. The other interesting regime is $|A| >1$ corresponding to the cases of Fig. 4,
we find the propagation speed of the wavefront is independent on the value of the strength, and remains about $0.24v_F$.
The physical reason for the unchanged speed can be explained as follows. For the regime of $|A| > 1$, the shock wave is
formed by colliding two initial separated clouds. As these two clouds are close each other and overlap gradually at the center of the trap,
the dispersive effect results in the formation of soliton trains. Due to the transverse instability,
the soliton trains quickly decay into vortex rings and the density bulge peak is formed by filling with the numerous vortex rings.
We calculated the propagation speed of the shock waves from the movements of self-steepen edges of the bulge peak. Therefore,
the obtained speeds actually correspond to the same expansion speed of the proliferation of the vortex rings.

In the right inset, we show the numerical results for the experimental situation $|A| = 29.2$
calculated from two different situations. One denoted by open square is obtained from the density profiles after evolving in the
harmonic trap (see Fig.\! 4(d)),
while the other by filled square is calculated from the density profiles after expansion for 1.5 \!ms after evolving in the harmonic trap
as performed in the experiment (see Fig.\! 1).
The speed of $0.302v_F$ obtained
from the profiles after the additional expansion is much larger than $0.243v_F$ calculated in the trap.
This is because such expansion method enhances the difference of the wavefront positions at two sequence times.
The result ($0.25v_F$, triangle) based on the DFT \cite{bul2012} and the experimental
data ($0.33v_F$, diamond) \cite{jos2011}
are also shown for comparison.

%%%%%%%%%%%%%%%%%%%%%%%%%%%%%%%%%%%%%%%%%%%%%%%%%%%%%%%%%%%%%%%%%
%%%%%%%%%%%%%%%%%%%%%%%%%%%%%%%%%%%%%%%%%%%%%%%%%%%%%%%%%%%%%%%%%

%%%%%%%%%%%%%%%%%%%%%%%%%%%%%%%%%%%%%%%%%%%%%%%%%%%%%%%%%%%%%%%%%
\section{Conclusion}
In conclusion, based on the order-parameter equation we have presented a detailed study for the
experiment on dynamics of shock waves in the unitary superfluid Fermi gas performed by the Thomas's group,
which shows good agreement with the experiment. The shock wave is studied by colliding two spatially
separated clouds. As the two clouds gradually overlap at the center of the trap, a soliton train forms
due to the dispersive effect provided by the quantum pressure term.
Due to the transverse instability, the soliton train decays very quickly into a large number of vortex rings.
The boxlike-shaped density peak observed in the experiment is then interpreted by the proliferation of the
vortex rings.  In addition, we have studied the mechanism of transition
from a sound wave to shock waves by calculating the speeds of the wavefronts and given an explanation
why the speed of the shock wave observed in the experiment is so close to the sound speed.
For a very small strength of the initial repulsive potential, sound wave is created
with the propagation speed of $0.347v_F$.  As the strength increases in the relatively
weak regime, the nonlinear effect leads to the formations of shock waves, and the speed decreases
below the sound speed as a scaling behavior. When the strength is moderate, the transversal dynamics
takes an effect to suppress the decrease of the speed.  For the large strength regime where the shock wave
is formed by colliding two separated clouds, the propagation speed of the wavefront is independent on the strength and
remains about $0.24v_F$. We understand this speed as the expansion speed of the proliferation of the vortex rings.
Finally, our numerical simulation demonstrates that the expansion image method results in calculating
the propagation speed from $0.243v_F$ to $0.302v_F$, which is very close to $0.33v_F$ of the experimental observation.

%%%%%%%%%%%%%

\acknowledgments
We would like to thank John Thomas and James Joseph for suggesting to study the propagation
speed of the shock wave fronts, and useful comments on the manuscript.
We also thank Han Pu for the critical comments on the manuscript.
W.W. thanks the hospitality of KITPC where part of the writing is done,
and useful discussions with Hui Zhai.  This work is supported by the NSFC under Grant (No. 11105039 and No. 11274092),
and Fundamental Research Funds for the Central Universities of China (Program No. 2012B05714 and No. 2014B11414).

\end{CJK*}

\begin{thebibliography}{10}
%review book
\bibitem{zwe2012} W. Zwerger, {ed.}, {\it The BCS-BEC Crossover and the Unitary
Fermi Gas}, Lecture Notes in Physics, Vol. 836 (Springer-Verlag, Berlin, 2012).

\bibitem{ket2008} W. Ketterle and M. W. Zwierlein, {\it Riv. Nuovo Cimento} \textbf{31}, 247 (2008).

\bibitem{all2012} A. Adams, L. D. Carr, T. Sch\"{a}fer, P. Steinberg and J. E. Thomas, New J. Phys. \textbf{14}, 115009 (2012).

%collective modes and sound
\bibitem{kin2004} J. Kinast, S. L. Hemmer, M. E. Gehm, A. Turlapov, and J. E. Thomas, Phys. Rev. Lett. \textbf{92}, 150402 (2004).

\bibitem{bar2004} M. Bartenstein, A. Altmeyer, S. Riedl, S. Jochim, C. Chin, J. H. Denschlag, and R. Grimm, Phys. Rev. Lett. \textbf{92},
203201 (2004); A. Altmeyer, S. Riedl, C. Kohstall, M. J. Wright, R. Geursen, M. Bartenstein, C. Chin, J. H. Denschlag, and R. Grimm,
{\it ibid.} \textbf{98}, 040401 (2007); A. Altmeyer, S. Riedl, M. J. Wright, C. Kohstall, J. H. Denschlag, and R. Grimm, Phys. Rev. A \textbf{76},
033610 (2007).

\bibitem{tey2013} M. K. Tey, L. A. Sidorenkov, Edmundo R. S\'{a}nchez Guajardo, R. Grimm, M. J. H. Ku, M. W. Zwierlein, Y. H. Hou, L. Pitaevskii and
S. Stringari, Phys. Rev. Lett. \textbf{110}, 055303 (2013).

\bibitem{jos2007} J. Joseph, B. Clancy, L. Luo, J. Kinast, A. Turlapov, and J. E. Thomas, Phys. Rev. Lett. \textbf{98}, 170401 (2007).

\bibitem{sid2013} L. A. Sidorenkov, M. K. Tey, R. Grimm, Y. -H. Hou, L. Pitaevskii and S. Stringari, Nature \textbf{498}, 78 (2013).

%vortice and soliton
\bibitem{zwi2005} M. W. Zwierlein, J. R. Abo-Shaeer, A. Schirotzek, C. H. Schunck, and W. Ketterle, Nature \textbf{435}, 1047 (2005).

\bibitem{yef2013} T. Yefsah, A. T. Sommer, M. J. H. Ku, L. W. Cheuk, W. Ji, W. S. Bakr and M. W. Zwierlein,  Nature \textbf{499}, 426 (2013).

\bibitem{ku2014} M. J. H. Ku, W. Ji, B. Mukherjee, E. Guardado-Sanchez, L. W. Cheuk, T. Yefsah, and M. W. Zwierlein, Phys. Rev. Lett.
\textbf{113}, 065301 (2014).

%shock wave in superfluid Fermi gases experimentally
\bibitem{jos2011} J. A. Joseph, J. E. Thomas, M. Kulkarni and A. G. Abanov, Phys. Rev. Lett. \textbf{106}, 150401 (2011).

%shock wave in classic fluid
\bibitem{whi1974} G. B. Whitman, {\it Linear and Nonlinear Waves}, (Wiley, New York, 1974).

\bibitem{lan1987} L. D. Landau and E. M. Lifshitz, {\it Fluid Mechanics}, (Pergamon Press, Loundon, 1987).

%shock wave in atomic BEC
%theory
\bibitem{kul2003}I. Kulikov and M. Zak, Phys. Rev. A \textbf{67}, 063605 (2003).

\bibitem{dam2004}B. Damski, Phys. Rev. A \textbf{69}, 043610 (2004).

\bibitem{kam2004}A. M. Kamchatnov, A. Gammal, and R. A. Kraenkel, Phys. Rev. A  \textbf{69}, 063605 (2004).

\bibitem{vic2004}V. M. P\'{e}rez-Garc\'{\i}a, V. V. Konotop, and V. A. Brazhnyi, Phys. Rev. Lett. \textbf{92}, 220403 (2004).

\bibitem{hof2006}M. A. Hoefer, M. J. Ablowitz, I. Coddington, E. A. Cornell, P. Engels, and V. Schweikhard, Phys. Rev. A \textbf{74}, 023623 (2006).

%experiments

\bibitem{cha2008} J. J. Chang, P. Engels, and M. A. Hoefer, Phys. Rev. Lett. \textbf{101}, 170404 (2008).

\bibitem{dut2001}Z. Dutton, M. Budde, C. Slowe and L. V. Hau, Science \textbf{293}, 663 (2001).

\bibitem{mep2009}R. Meppelink, S. B. Koller, J. M. Vogels, P. van der Straten, E. D. van Ooijen, N. R. Heckenberg, H. Rubinsztein-Dunlop,
S. A. Haine and M. J. Davis, Phys. Rev. A \textbf{80}, 043606 (2009).


%shock wave in Fermi gas theoretically

%{dispersive effects in the Fermi gas}
\bibitem{bul2012} A. Bulgac, Y.-L. Luo, and K. J. Roche, Phys. Rev. Lett. \textbf{108}, 150401 (2012).

\bibitem{anc2012} F. Ancilotto, L. Salasnich, and F. Toigo, Phys. Rev. A \textbf{85}, 063612 (2012).

\bibitem{sal2011} L. Salasnich, Europhys. Lett. \textbf{96}, 40007 (2011).

\bibitem{anc2013} F. Ancilotto, L. Salasnich, and F. Toigo, J. Low Temp. Phys. \textbf{171}, 329 (2013).

\bibitem{lsa2013} L. Salasnich, Few-Body Syst. \textbf{54}, 697 (2013).

%viscosity in UFG
\bibitem{son2007} D. T. Son, Phys. Rev. Lett. \textbf{98}, 020604 (2007).

\bibitem{cao2011} C. Cao, E. Elliott, H. Wu and J. E. Thomas, New J. Phys. \textbf{13}, 075007 (2011);
                  J. A. Joseph, E. Elliott, and J. E. Thomas, arXiv:1410.4835.

\bibitem{wal2012} G. Wlaz{\l}owski, P. Magierski, and J. E. Drut, Phys. Rev. Lett. \textbf{109}, 020406 (2012).

%speed of shock wave front
\bibitem{low2013} N. K. Lowman and M. A. Hoefer, Phys. Rev. A \textbf{88}, 013605 (2013).

%order parameter equation
\bibitem{sal2008} L. Salasnich, N. Manini, and F. Toigo, Phys. Rev. A \textbf{77}, 043609 (2008).

\bibitem{adh2008} S. K. Adhikari, Phys. Rev. A \textbf{77}, 045602 (2008).

\bibitem{sal12008} L. Salasnich and F. Toigo, Phys. Rev. A \textbf{78}, 053626 (2008).

\bibitem{adh12008} S. K. Adhikari and L. Salasnich, Phys. Rev. A \textbf{78}, 043616 (2008).

\bibitem{wen2008} W. Wen, Y. Zhou, and G. Huang, Phys. Rev. A \textbf{77}, 033623 (2008).

%equation of state
\bibitem{ast2004} J. Carlson, S.-Y. Chang, V. R. Pandharipande, and K. E. Schmidt, Phys. Rev. Lett. \textbf{91}, 050401 (2003).

\bibitem{bul2006} A. Bulgac, J. E. Drut and P. Magierski, Phys. Rev. Lett. \textbf{96}, 090404 (2006).

%Bertsch parameter
\bibitem{nav2010} N. Navon, S. Nascimb\`{e}ne, F. Chevy, and C. Salomon, Science \textbf{328}, 729 (2010).

\bibitem{ku2012} M. J. H. Ku, A. T. Sommer, L. W. Cheuk, and M. W. Zwierlein, Science \textbf{335}, 563 (2012).

%quantum pressure
\bibitem{zub2009} A. L. Zubarev, J. Phys. B: At. Mol. Opt. Phys. \textbf{42}, 011001 (2009).

\bibitem{zub12009} A. L. Zubarev and M. Zoubarev, Europhys. Lett. \textbf{87}, 33001 (2009).

\bibitem{and2010} A. Csord\'{a}s, O. Alm\'{a}sy, and P. Sz\'{e}pfalusy, Phys. Rev. A \textbf{82}, 063609 (2010).

%hydrodynamics equations
\bibitem{men2002} C. Menotti, P. Pedri, and S. Stringari, Phys. Rev. Lett. \textbf{89}, 250402 (2002).

%old order parameter equation
\bibitem{kim2004} Y. E. Kim and A. L. Zubarev, Phys. Rev. A \textbf{70}, 033612 (2004).

%time-dependent density functional theory
\bibitem{bulg2013} A. Bulgac and M. M. Forbes, in {\it Quantum Gases: Finite Temperature and Non-Equilibrium
Dynamics}, Cold Atoms Series, Vol. 1, edited by N. P. Proukakis {\it et al}. (Imperial College
Press, London, 2013), Chap. 26.

\bibitem{bulg2011} A. Bulgac, Y.-L. Luo, P. Magierski, K. J. Roche, and Y. Yu, Science \textbf{332}, 1288 (2011).

\bibitem{bul2014} A. Bulgac, Michael McNeil Forbes, Michelle M. Kelley, K. J. Roche, and G. Wlaz{\l}owski,  Phys. Rev. Lett. \textbf{112}, 025301 (2014).

\bibitem{wla2014} G. Wlaz{\l}owski, A. Bulgac, M. M. Forbes, and K. J. Roche, arXiv:1404.1038.

\bibitem{forb2013} Michael McNeil Forbes and R. Sharma, Phys. Rev. A \textbf{90}, 043638 (2014).


%numerical method
\bibitem{adh12010} S. K. Adhikari, J. Phys. B: At. Mol. Opt. Phys. \textbf{43}, 085304 (2010).

% Fermi gas instability
\bibitem{cet2013} A. Cetoli, J. Brand, R. G. Scott, F. Dalfovo, and L. P. Pitaevskii,
Phys. Rev. A \textbf{88}, 043639 (2013).

\bibitem{mun2014} A. M. Mateo and J. Brand, Phys. Rev. Lett. \textbf{113}, 255302 (2014).

\bibitem{wen2013} W. Wen and H.-J. Li, J. Phys. B: At. Mol. Opt. Phys. \textbf{46}, 035302 (2013).

%soliton-trains and transverse instability
\bibitem{fed2000} D. L. Feder, M. S. Pindzola, L. A. Collins, B. I. Schneider, and C. W. Clark,
Phys. Rev. A \textbf{62}, 053606 (2000).

\bibitem{hoe2009} M. A. Hoefer and B. Ilan, Multiscale Model. Simul. \textbf{10}, 306 (2012).
%sound speed
\bibitem{cap2006} P. Capuzzi, P. Vignolo, F. Federici, and M. P. Tosi, Phys. Rev. A \textbf{73},
021603(R) (2006).

\bibitem{wen2010} W. Wen, S. -Q. Shen and G. Huang, Phys. Rev. B \textbf{81}, 014528 (2010).

\bibitem{hei2006} H. Heiselberg, Phys. Rev. A \textbf{73}, 013607 (2006).

\bibitem{dam12004} B. Damski,  J. Phys. B: At. Mol. Opt. Phys. \textbf{37}, L85 (2004).

\end{thebibliography}
\end{document}